\documentstyle[aps,prl,twocolumn, epsf]{revtex}
\def\be{\begin{equation}}
\def\ee{\end{equation}}
\def\bea{\begin{eqnarray}}
\def\eea{\end{eqnarray}}
\def\bma{\begin{mathletters}}
\def\ema{\end{mathletters}}

\def\0{\overline{0}}
\def\p{\overline{\pi}}
\def\q0{\underline{0}}
\def\qp{\underline{\pi}}
\def\H{{\cal H}}

\def\C{{\cal C}}

\def\R{{\cal R}}

\newcommand{\bra}[1]{\mbox{$\langle #1 |$}}
\newcommand{\ket}[1]{\mbox{$| #1 \rangle$}}
\newcommand{\braket}[2]{\mbox{$\langle #1  | #2 \rangle$}}
\newcommand{\proj}[1]{\ket{#1}\!\bra{#1}}

\begin{document}
         
\draft

\title{Storing quantum dynamics in quantum states: \\
 stochastic programmable gate for U(1) operations}

\author{G. Vidal$^1$, L. Masanes$^{1,2}$ and J. I. Cirac$^1$}

\address
{$^1$ Institut f\"ur Theoretische Physik, Universit\"at Innsbruck,A-6020 Innsbruck, Austria\\
$^2$ Departament d'Estructura i Constituents de la Mat\`eria, Universitat de Barcelona, E-08028 Barcelona, Spain.}

\date{\today}

\maketitle

\begin{abstract}
We show how quantum dynamics can be captured in the state of a quantum system, in such a way that the system can be used to stochastically perform, at a later time, the stored transformation perfectly on some other quantum system. Thus programmable quantum gates for quantum information processing are feasible if some probability of failure ---that we show to decrease exponentially with the size of the storing resources---  is allowed. 

\end{abstract}

\pacs{PACS Nos. 03.67.-a, 03.65.Bz}

\bigskip

Quantum Information Science investigates the potential of Quantum Mechanics to process and transmit information in novel ways. Quantum systems are usually conceived as containers for data, which is processed by means of a unitary evolution.
In this Letter we will explore to which extend Quantum Mechanics allow for the processing itself ---i.e. instead of the data--- to be stored in a quantum system. In particular, we will present a scheme to encode unitary transformations in, and to stochastically retrieve them from, quantum states. The practical importance of this result relies on the fact that, once the operation has been captured in a quantum state, it can be processed by means of any standard state manipulation technique. And thus, for instance, the operation can be simply kept for later use, but it can also be transmited to a remote party (e.g., using teleportation) or can by estimated by means of a proper measurement.

 The storage of operations is, as explained below, necessarily imperfect. Our scheme will fail with a probability $\epsilon$ that exponentially decreases with the number of qubits in which the operation has been encoded. More specifically, we will show how to store, using $N$ qubits and with probability $\epsilon=2^{-N}$ of failure in its later retrieval, an arbitrary rotation of a qubit around the $\hat{z}$ axis.
For $N=1$ we will prove that our scheme is optimal, i.e., it has the minimal error probability ever possible, whereas for $N>1$ several evidence in the same direction will be presented.

Let us start by considering two quantum systems, that we will call {\em program} and {\em data registers}, with corresponding Hilbert spaces $\H_P$ and $\H_D$. A {\em program state} $\ket{U} \in \H_P$ will be said to store the transformation $U$, if some ``fixed'' protocol employing $\ket{U}$ is able to perform $U$ on an arbitrary {\em data state} $\ket{d}\in \H_D$. Here, a ``fixed'' protocol means that the manipulation of the joint state
\be
\ket{d}\otimes\ket{U} 
\label{estat}
\ee
does not require knowing the operation $U$ nor the data $\ket{d}$. A device able to transform state (\ref{estat}) into 
\be
U\ket{d}\otimes \ket{\R_{d,U}},
\label{universal}
\ee
where $\ket{\R_{d,U}}$ is just some residual state, is known as a {\em programmable quantum gate} \cite{Nie97}. Thus, in a similar fashion as most ``classical'' computers take both program and data as input bit strings, a programmable or universal quantum gate is a device whose action $U$ on an arbitrary data state $\ket{d}$ is completely determined by the program state $\ket{U}$.

 Nielsen and Chuang analyzed in Ref. \cite{Nie97} the possibility of constructing one such gate. Its total dynamics are described in terms of a fixed unitary operator G,
\be
G[\ket{d}\otimes\ket{U}] = (U\ket{d})\otimes\ket{\R_U},
\label{perfect}
\ee
where the residual state $\ket{\R_U}$ was showed to be independent of $\ket{d}$. Also the following important result was proved: any two inequivalent operations $U$ and $V$ require orthogonal program states, that is $\braket{U}{V}=0$, if the same transformation $G$ is to implement them according to Eq. (\ref{perfect}).
This means that in order to perfectly store one operation $U_i$, chosen from a finite set $\{U_i\}_{i\in I}$, a vector state $\ket{U_i}$ belonging to an orthonormal basis $\{\ket{U_i}\in\H_P\}_{i\in I}$ has to be used. In other words, different operations of the gate necessarily correspond to mutually {\em distinguishable} programs. This has two direct implications. First, a classical binary string could have been used in the first place as a program (there is no gain in using quantum states for this purpose). The second consequence concerns the feasibility of such gates: even for the simplest data register, a qubit (i.e. $\H_D=\C^2$), the set of unitary transformations, $SU(2)$, is infinite. Therefore no universal gate implementing an arbitrary (say) one-qubit operation can be constructed using a program register whose Hilbert space $\H_P$ has finite dimension. 

 Here we will assume, nevertheless, that only $N$ qubits are available as a program register, and thus $\H_P=\C^{2 \otimes N}$ is finite dimensional. For simplicity, we will restrict our attention to one-qubit operations of the form
\be
U_{\alpha} \equiv \exp(i\alpha \frac{\sigma_z}{2}),
\label{unialpha}
\ee
for an arbitrary angle $\alpha\in [0,2\pi)$, which correspond to arbitrary rotations around the $\hat{z}$ axis of a spin $1/2$ particle \cite{comment}. We would then like to answer the question: To what extend can $N$ qubits store an arbitrary operation $U_{\alpha}$?  

The quality of the storage is determined by how well the operation can be retrieved, that is, by how well it can be finally performed on the unknown data state $\ket{d}$. One possibility would be to consider {\em approximate} transformations, with the output state of the gate being an approximation to $U_{\alpha}\ket{d}$. But this can already be achieved by classically encoding a truncated binary expansion of the angle $\alpha$\cite{approx}. Alternatively, as we will next discuss, {\em stochastic} transformations may be considered. In this case the programmable gate does not always succeed at performing $U$ after processing the program $\ket{U}$, but when it does succeed, then the output state is exactly $U_{\alpha}\ket{d}$. Of course, we also want to be able to know whether the gate achieved its goal or not. Reasonably, the a priori probability of success is a good figure of merits for this kind of programmable gates. Since in principle such probability $p_{\alpha}^d$ may depend both on the data $\ket{d}$ and on the operation $U_{\alpha}$ under consideration, we will use its average 
\be
\langle p\rangle \equiv \int_{C^2} d(d) \int \frac{d{\alpha}}{2\pi} ~~p_{\alpha}^{d}
\label{aveprob}
\ee
to quantify the performance of the gate.

\begin{figure}
 \epsfysize=1.6cm
\begin{center}
 \epsffile{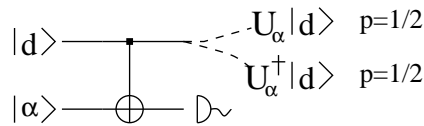}
\end{center}
 \caption{Optimal stochastic quantum programmable gate with a single-qubit program register. Data and program states $\ket{d}$ and $\ket{\alpha}$ are transformed, depending on the result of a measurement on the program register, either into $U_{\alpha}\ket{d}$ or $U_{\alpha}^{\dagger}\ket{d}$, with error probability $\epsilon = 1/2$.  \label{fig1}}
\end{figure}

Let us suppose, first, that only one qubit, i.e. $N=1$, is available to encode any of the transformations $U_{\alpha}$. In this case the {\em equatorial state}
\be
\ket{\alpha} \equiv \frac{1}{\sqrt{2}} (e^{i\alpha/2}\ket{0}+e^{-i\alpha/2}\ket{1})
\label{equatorial}
\ee
can be used to store  $U_{\alpha}$, in the sense that a CNOT gate,
$\proj{0}\otimes I + \proj{1}\otimes \sigma_x$,
---taking the data and program register as control and target qubits, respectively--- will be able to transform the data state $\ket{d}$ according to $U_{\alpha}$, with probability $1/2$, for all $\ket{d}$ and all $U_{\alpha}$ (see FIG. \ref{fig1}). Indeed, it is straightforward to check that
\be
\ket{d}\otimes\ket{\alpha} \stackrel{\mbox{C-NOT}}{\longrightarrow} \frac{1}{\sqrt{2}}(U_{\alpha}\ket{d}\otimes\ket{0} + U_{\alpha}^{\dagger}\ket{d}\otimes \ket{1}),
\label{optim1}
\ee
 and therefore a projective measurement in the $\{\ket{0}, \ket{1}\}$ basis of the program register will make the data qubit collapse either into the desired state $U_{\alpha}\ket{d}$ or into the wrong state $U_{\alpha}^{\dagger}\ket{d}$, with the announced probabilities.

 In order to see that no scheme exists better than the one above, let us consider the most general stochastic programmable gate using a single qubit as a program register.
It can always be represented by a unitary transformation $G_{s}$ given by
\bea
G_s[\ket{d}\otimes\ket{U_{\alpha}}\otimes\ket{0}] &\equiv& \sqrt{p^d_{\alpha}} (U_{\alpha}\ket{d})\otimes\ket{\tau_{\alpha}^{d}} \nonumber \\
&+& \sqrt{1-p^d_{\alpha}} \ket{\chi^d_{\alpha}},
\label{Gs}
\eea
taking the data and program states, together with a fixed state $\ket{0}$ of a third (ancillary) system $\H_A$, into $U_{\alpha}\ket{d}$ with probability $p_{\alpha}^d$. Note that all kets appearing in Eq. (\ref{Gs}) are normalized vectors.
We demand that for all possible $d,d', \alpha,\alpha'$, the state $\braket{\tau^d_{\alpha}}{\chi_{\alpha'}^{d'}} \in \H_D$ vanishes. This is equivalent to requiring that by means of a measurement ---onto the support $\Pi_{\tau} \subseteq \H_P\otimes\H_A$ of the vectors $\ket{\tau_{\alpha}^d}$ and its complementary subspace $\Pi_{\tau}^{\perp}$--- we are able to know whether the gate succeeded or not.

Since $G_s$ is a linear transformation, by decomposing $\ket{d}$ as $a\ket{\0}+b\ket{\p}$, where $a,b$ are complex coefficients ($|a|^2+|b|^2=1$) and $\ket{\0}\equiv (\ket{0}+ \ket{1})/\sqrt{2}$, $\ket{\p }\equiv i(\ket{0}- \ket{1})/\sqrt{2}$, we obtain that the RHS of Eq. (\ref{Gs}) must be equal to
\bea
a [\sqrt{p^{\0}_{\alpha}} (U_{\alpha}\ket{\0})\otimes\ket{\tau_{\alpha}^{\0}}+ \sqrt{1-p^{\0}_{\alpha}} \ket{\chi^{\0}_{\alpha}}], \nonumber \\
+b [\sqrt{p^{\p}_{\alpha}} (U_{\alpha}\ket{\p})\otimes\ket{\tau_{\alpha}^{\p}}+ \sqrt{1-p^{\p}_{\alpha}} \ket{\chi^{\p}_{\alpha}}].
\eea
This implies that the probability of success $p_{\alpha}^d$ and the vector $\ket{\tau_{\alpha}^d}$, from now on $p_{\alpha}$ and $\ket{\tau_{\alpha}}$, do not depend on the data $\ket{d}$.
%, and that $\ket{\chi_{\alpha}^d} = a\ket{\chi^{\0}_{\alpha}} + b \ket{\chi^{\p}_{\alpha}}$
 On the other hand, the most general codification scheme of $U_{\alpha}$ on a qubit, $[0,2\pi) \rightarrow \C^2$, can be parameterized as $\ket{U_{\alpha}} \equiv A(\alpha)\ket{\q0} + B(\alpha)\ket{\qp}$, where $A(\alpha)$ and $B(\alpha)$ are complex functions ($\braket{U_{\alpha}}{U_{\alpha}}=|A(\alpha)|^2+|B(\alpha)|^2 + 2\mbox{Re}[A(\alpha)B^*(\alpha)\braket{\q0}{\qp}] =1$) and the states $\ket{\q0}$ and $\ket{\qp}$ correspond to the (not necessarily orthonormal) programs $\ket{U_0}$ and $\ket{U_{\pi}}$. Expanding now $\ket{U_{\alpha}}$ in Eq. (\ref{Gs}) we find that its RHS must read
\bea
&A&(\alpha) [\sqrt{p_0} U_0\ket{d} \otimes\ket{\tau_{0}} + \sqrt{1-p_0}\ket{\chi^d_0}] \nonumber\\
+&B&(\alpha) [\sqrt{p_{\pi}} U_{\pi}\ket{d} \otimes\ket{\tau_{\pi}} + \sqrt{1-p_{\pi}}\ket{\chi^d_{\pi}}]
\eea 
for any $\ket{d}$, which readily implies that the states $\ket{\tau_{\alpha}} (\equiv \ket{\tau})$ do not depend on $\alpha$ and that $\sqrt{p_{\alpha}} U_{\alpha} = A(\alpha)\sqrt{p_{0}} U_0 + B(\alpha)\sqrt{p_{\pi}} U_{\pi}$. This last equation leads to $A(\alpha) = \sqrt{p_{\alpha}/p_{0}}\cos(\alpha/2)$ and $B(\alpha) = \sqrt{p_{\alpha}/p_{\pi}}\sin(\alpha/2)$. If we now substitute these in state $\ket{U_{\alpha}}$, from its normalization we obtain 
\be
p_{\alpha} = (\frac{\cos^2 \frac{\alpha}{2}}{p_{0}} + \frac{\sin^2 \frac{\alpha}{2}}{p_{\pi}} + 2\frac{\cos \frac{\alpha}{2}\sin \frac{\alpha}{2}\mbox{Re}[\braket{\q0}{\qp}]}{\sqrt{p_{0}p_{\pi}}})^{-1}. \nonumber
\ee
Recall that our goal is to maximize the average probability of success (\ref{aveprob}). Without loss of generality we can require that $p_{0} \geq p_{\alpha}$ \cite{explicacio}, which corresponds to choosing $\mbox{Re}[\braket{\q0}{\qp}]=0$. It is now easy to compute $\langle p\rangle$, which reads $\sqrt{p_0p_{\pi}}$. Substituting all the previous findings in Eq. (\ref{Gs}), and computing the scalar product of $G_s[\ket{\0}\otimes\ket{\q0}\otimes \ket{0}]$ and $G_s[\ket{\p}\otimes\ket{\qp}\otimes\ket{0}]$ we obtain
\be
0 = -\sqrt{p_0p_{\pi}}  + \sqrt{1-p_0}\sqrt{1-p_{\pi}}\braket{\chi_0^{\0}}{\chi_{\pi}^{\p}}. 
\ee
That is, $\sqrt{p_0p_{\pi}}$ is at most $\sqrt{1-p_0}\sqrt{1-p_{\pi}}$. The most favorable case corresponds to $p_{0} = 1- p_{\pi}$, and therefore the maximal $\langle p\rangle = \sqrt{p_0 p_{\pi}}$ is $1/2$, achieved when $p_{\alpha}=1/2$ is constant. This ends the proof that Eqs. (\ref{equatorial}) and (\ref{optim1}) constitute the optimal protocol for storing and stochastically retrieving an operation $U_{\alpha}$ in a single qubit, with the associated error $\epsilon \equiv 1- \langle p \rangle$ being $1/2$.

\begin{figure}
 \epsfysize=2.0cm
\begin{center}
 \epsffile{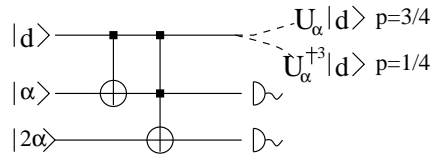}
\end{center}
 \caption{ The gate of FIG. \ref{fig1} can be improved by making a conditional correction of the output after its C-NOT gate. This is achieved by means of a Toffoli gate, which acts as a C-NOT between the first and third line of the circuit only when the second line carries a $\ket{1}$, corresponding to a failure in FIG.  \ref{fig1}. A measurement on the program qubits in the $\{\ket{0},\ket{1}\}$ basis will reveal whether the gate failed (this happens when outcome $1$ is obtained from both registers, i.e. $\epsilon = 1/4$).
 \label{fig2}}
\end{figure}

 We now move to consider the storage of $U_{\alpha}$ using more qubits, $N>1$. When the previous scheme fails, not only has the data $\ket{d}$ not yet been processed properly, but in addition it has been modified in an unwished manner (which is unknown to the user of the gate) into $U_{\alpha}^{\dagger}\ket{d}$. However, a single second go of the previous gate may correct $U_{\alpha}^{\dagger}\ket{d}$ into $U_{\alpha}\ket{d}$ at once. This is achieved by just inserting $U_{\alpha}^{\dagger}\ket{d}$ in the gate of Eq. (\ref{optim1}), together with a new program state, namely $\ket{2\alpha}$ (see FIG. \ref{fig2}). That, is, the two-qubit program $\ket{\alpha}\otimes\ket{2\alpha}$ stores $U_{\alpha}$ with a probability of failure $\epsilon=1/4$ in the retrieval stage.

\begin{figure}
 \epsfysize=3cm
\begin{center}
 \epsffile{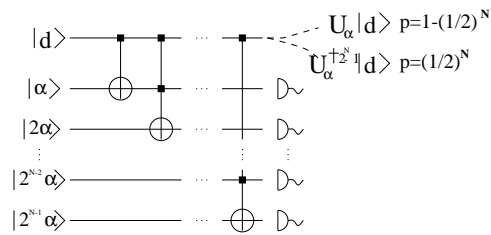}
\end{center}
 \caption{Stochastic programmable quantum gate with a $N$-qubit program register and success probability $p = 1-(1/2)^{N}$, i.e. $\epsilon = (1/2)^{N}$. 
%If no final measurement on the program register is made, or its result is ignored, then this circuit can be regarded as an approximate programmable quantum gate with performance fidelity $F\geq 1-(1/2)^{N}$. 
The gate only fails when all the outcomes of a $\{\ket{0},\ket{1}\}$-basis measurement on the $N$ register qubits are $1$.
\label{fig3}}
\end{figure}

 In case of a new failure, the state of the system becomes $U_{\alpha}^{\dagger 3}\ket{d}$. We can insert again this state, together with state $\ket{4\alpha}$, into the elementary gate. If we  keep on obtaining failures, we can try to correct the state as many times as wished, provided that the state $\ket{2^{l-1}\alpha}$ is available at the $l$th attempt. Therefore, for any $N$, the $N$-qubit state 
\be
\ket{U_{\alpha}^N}\equiv\bigotimes_{l=1}^{N}\ket{2^{l-1}\alpha}
\label{superequ}
\ee
can be used to implement the transformation $U_{\alpha}$ with probability $1- (1/2)^N$ \cite{Cirac00}.
 The corresponding stochastic programmable gate (see FIG. \ref{fig3}), consists of the unitary transformation of $\ket{d}\otimes\ket{U^N_{\alpha}}$ into 
\be
\frac{1}{2^{N/2}}(\sqrt{2^N\!-\!1}~U_{\alpha}\ket{d}\otimes\ket{\tau} + U_{\alpha}^{(2^N\!-\!1)\dagger}\ket{d}\otimes \ket{\chi})
\label{optimN}
\ee
and of a posterior measurement of the program register (either in state $\ket{\tau}$ or $\ket{\chi}\equiv \ket{1}^{\otimes N}$, $\braket{\tau}{\chi}=0$). Its failure probability, $\epsilon = (1/2)^{N}$, decreases exponentially with the size $N$ of the program register. 

 We are tempted to conjecture that, for any $N$, Eqs. (\ref{superequ})-(\ref{optimN}) define again an optimal protocol to store and stochastically retrieve $U_{\alpha}$. Notice, on the one hand, that the $N$-qubit unknown state $\ket{U^N_{\alpha}}$ has maximal entropy, since $\int d\alpha/(2\pi) \proj{U^N_{\alpha}}=(I/2)^{\otimes N}$, where $I$ is the identity operator in $\C^2$. That is, this program state carries as much information as possible, with $N$ bits of information about $\alpha$ being extractable from it for large $N$ \cite{Lluis}. On the other hand, we will now prove that our scheme is the optimal way of retrieving $U_{\alpha}$ from the program $\ket{U^N_{\alpha}}$ as given in Eq. (\ref{superequ}). 

Indeed, let $G_s^N$ be a unitary transformation producing $U_{\alpha}\ket{d}$ from $\ket{d}\otimes\ket{U^N_{\alpha}}$, with probability $p_{\alpha}$ (we already learned, from the single-qubit case, that the probability of success is independent of the data state $\ket{d}$). From $G^N_s$ we can construct another gate $G_{s'}^{N}$ with constant probability of success $p'_{\alpha} = \langle p\rangle_{G^N_s}$, where  $\langle p\rangle_{G^N_s} \equiv \int d\alpha/(2\pi) p_{\alpha}$ is the average probability of success of $G^N_s$, precisely the quantity to be maximized. The construction goes as follows. Given a program state $\ket{U_{\alpha}^N}$, we will randomly choose an angle $\alpha_0 \in [0,2\pi)$ and will transform the program into $\ket{U_{\alpha+\alpha_0}^N}$. This can be achieved by performing $U_{\alpha_0}\otimes U_{\alpha_0}^2\otimes...\otimes U_{\alpha_0}^{2^{N-1}}$ on $\ket{U_{\alpha}^N}$. Then we will run $G_s^N$ on $\ket{d}$ using the new program, to obtain $U_{\alpha+\alpha_0}\ket{d}$ with probability $p_{\alpha+\alpha_0}$. Finally, we will perform $U_{\alpha_0}^{\dagger}$ on $U_{\alpha+\alpha_0}\ket{d}$. The overall effect is the promised gate $G_{s'}^N$, and therefore we only need to optimize over programmable gates with constant success probability $p$,
\be
\ket{d}\otimes\ket{U_{\alpha}^N}\otimes\ket{0} \rightarrow \sqrt{p}U_{\alpha}\ket{d}\otimes \ket{\tau_{\alpha}} + \sqrt{1-p}\ket{\chi_{\alpha}^d}.
\ee
Let us choose the data $\ket{d}$ to be an equatorial state $\ket{\beta}$ of angle $\beta$, so that $U_{\alpha}\ket{\beta}=\ket{\alpha+\beta}$. Unitarity of the whole transformation $G_s'$ implies, if $\beta'\equiv \pi + \beta + \alpha - \alpha'$, that
\label{product}
\be
\braket{\beta'}{\beta}\braket{U^N_{\alpha'}}{U^N_{\alpha}} = (1-p)\braket{\chi_{\alpha'}^{\beta'}}{\chi_{\alpha}^{\beta}}
\label{product}
\ee
(notice that $\bra{\beta'}U_{\alpha'}^{\dagger}U_{\alpha}\ket{\beta} = 0$). The absolute value of the LHS of Eq. (\ref{product}) is now at most $1-p$. But we can easily compute the above scalar products using Eq. (\ref{equatorial}), to get the bound $(\sin 2^{N-1}(\alpha-\alpha'))/2^N \leq 1-p.$
Finally, taking $\alpha-\alpha' = \pi (1/2)^{N}$ we get that $p \leq 1- (1/2)^{N}$, as we wanted to prove.
 Thus, once the operation $U_{\alpha}$ has been encoded in $N$ qubits as $\ket{U_{\alpha}^N}$, the optimal extraction protocol necessarily fails with probability $\epsilon = (1/2)^N$. However, whether our encoding is also optimal, remains as an open question for $N>1$.

 Notice that if we want to warrant a priori a successful implementation of $U_{\alpha}$, infinitely many qubits are required for the program register, as originally stated in \cite{Nie97}. Interestingly enough, the average length of the program required to perform $U_{\alpha}$ with certainty is, in contrast, very small. Indeed, since with probability $p_1=1/2$ the gate of Eq. (\ref{optim1}) achieves the goal after using a single-qubit program; with probability $p_2=1/4$ a two-qubit program is sufficient; etc, the average length $\langle N \rangle$ of the required program is
\be
\langle N \rangle \equiv \sum_{N=1}^{\infty} p_N N = \sum_{N=1}^{\infty} \frac{N}{2^N} = 2.
\label{average}
\ee
That is, {\em on average}, a two-qubit program is sufficient to store and retrieve with certainty any operation $U_{\alpha}$. 

 Let us finally comment on how the storage of operations can be applied in the context of quantum remote control, as introduced by Huelga {\em et al.} in \cite{Hue00}. Suppose two distant parties, Alice and Bob, try to process some data state $\ket{d}$ of, say, a qubit, according to some unitary operation $U$. Alice possesses a device able to perform $U$, whereas Bob has the qubit in state $\ket{d}$. Their goal is that Bob ends up with the processed state $U\ket{d}$. 
 If the internal state of Alice's device cannot be teleported, then the optimal protocol \cite{Hue00} is to use standard teleportation \cite{tele} to send the data from Bob to Alice, who will use the device to process it and will teleport it back to Bob. 
This scheme requires two-way classical communication, and the coexistence in time and space of the data $\ket{d}$ and the device that performs $U$. 

If, alternatively, Alice codifies the operation $U$ in a quantum state using the scheme we have discussed, and then teleports the state to Bob, classical communication only from Alice to Bob is required to achieve quantum remote control. In addition Bob can receive the codified operation even when the data state $\ket{d}$ is not yet available. The price to be paid, however, is that the scheme only succeeds with some probability. 
Taking into account that a general $SU(2)$ operation decomposes into three rotations $U_{\alpha}$ \cite{comment}, each of these requiring, on average, a two-qubit program, and that teleportation of an equatorial state uses $1$ bit of communication and $1$ ebit of entanglement \cite{lo}, we conclude that on average $6$ ebits of entanglement have to be consumed and Alice has to send $6$ bits of communication to Bob in order to remotely perform a general $U\in SU(2)$.

Summarizing, we have presented a scheme for storing any unitary operation in a finite number of qubits, in a way that it can be stochastically retrieved at a later time. It would be interesting to know which are the minimal resources needed, per operation, in order to store and retrieve a large amount of them with asymptotic perfection. The results of D\"ur {\em et al} \cite{Wolf} represent a promising first step in this direction.

 We thank W. D\"ur for useful comments. This work was supported by the Austrian Science Foundation (SFB project 11), the Spanish newspaper ``La Vanguardia'', the Institute for Quantum Information GmbH and the European Community (HPMF-CT-1999-00200 and EQUIP project IST-1999-11053).

\begin{references}

\bibitem{Nie97} M.A. Nielsen and I.L. Chuang, Phys. Rev. Lett. {\bf 79} 321 (1997).
\bibitem{comment} Recall that any unitary transformation on a qubit can be obtained by composing three of such rotations and, for instance, two fixed $\pi/2$ rotations along the $\hat{y}$ axis, as in the Euler angles' construction. 

\bibitem{approx} Suppose that only $N=4$ qubits are available to store $U_{\alpha}$, and that $\alpha/2\pi=.0110101...$. Then we can store the first four digits, $0110$, classically. The gate will read them and perform $U_{\tilde{\alpha}}$, where $\tilde{\alpha}/2\pi= 0.0110$. The error in the fidelity of the resulting output tends to $0$ exponentially fast in $N$. 

\bibitem{explicacio} Given any stochastic programmable gate $G_s$ with success probability $p_{\alpha}$ for $U_{\alpha}$, we can construct another one, $G_s'$, from it with associated probability $p_{\alpha+\alpha_0}$ (and thus with the same average probability) as follows. $G_s'$ consists in first performing $U_{-\alpha_0}$ to the data $\ket{d}$ and then performing $G_s$ with the program $\ket{U_{\alpha+\alpha_0}}$, the probability of success being $p_{\alpha+\alpha_0}$. This program corresponds, therefore, to the program $\ket{U_{\alpha}'}$ of $G_s'$, with the modified success probability. Therefore we can always choose $p_0\geq p_{\alpha}$.


\bibitem{Cirac00} The several-step correction character of our scheme for implementing $U_{\alpha}$ is inspired in the one used by J. I. Cirac, W. D\"ur, B. Kraus and M. Lewenstein,  Phys. Rev. Lett. 86, 544 (2001) to implement a non-local unitary operation. In the present context all intermediate measurements and conditional actions can be substituted by a single unitary operation, as described in Figs. 2 and 3. In the text we have presented the several-measurement version for pedagogical reasons.


\bibitem{Lluis} L. Masanes et al, {\em in preparation}.

%\bibitem{acin}  A. Ac\'{\i}n, E. Jan\'e and G. Vidal, ``Optimal estimation of quantum dynamics'', quant-ph/0012015.

\bibitem{tele} C. Bennett, G. Brassard, C. Crepeau, R. Josza, A. Peres and W.K. Wootters, Phys. Rev. Lett. 70, 1895 (1993).

\bibitem{Hue00} S.F. Huelga, J.A. Vaccaro, A. Chefles and M.B. Plenio,  
quant-ph/0005061.

\bibitem{lo} See, e.g., H.-K. Lo, Phys. Rev. A 62, 012313 (2000).


\bibitem{Wolf} W. D\"ur and J. I. Cirac, quant-ph/0012148.

\end{references}
\end{document}